\documentclass[twocolumn,aps,prb,showpacs]{revtex4}
\usepackage{mathrsfs}
\usepackage{bm}
\usepackage{multirow}
\usepackage{amsmath,amsfonts,amssymb}
\usepackage{array,booktabs}
\usepackage{graphicx,times,graphics,color,epsfig}

\begin{document}

\title{Structures and Dielectric Properties of Amorphous High-$\kappa$ Oxides: HfO$_2$, ZrO$_2$ and their alloys}
\author{Yin Wang$^1$}
\email{yinwang@hku.hk}
\author{Ferdows Zahid$^1$}
\author{Jian Wang$^1$}
\author{Hong Guo$^{2, 1}$}
\affiliation{$^1$Department of Physics and the center of theoretical and
computational physics, The University of Hong Kong, Pokfulam Road, Hong Kong SAR, China \\
$^2$Center for the Physics of Materials and Department of Physics, McGill University, Montreal, PQ, Canada, H3A 2T8}

\begin{abstract}
High-$\kappa$ metal oxides are a class of materials playing an increasingly important role in modern device physics and technology. Here we report  theoretical investigations of the properties of structural and lattice dielectric constants of bulk amorphous metal oxides by a combined approach of classical molecular dynamics (MD) - for structure evolution, and quantum mechanical first principles density function theory (DFT) - for electronic structure analysis. Using classical MD based on the Born-Mayer-Buckingham potential function within a melt and quench scheme, amorphous structures of high-$\kappa$ metal oxides Hf$_{1-x}$Zr$_x$O$_2$ with different values of the concentration $x$, are generated. The coordination numbers and the radial distribution functions of the structures are in good agreement with the corresponding experimental data. We then calculate the lattice dielectric constants of the materials from quantum mechanical first principles, and the values averaged over an ensemble of samples agree well with the available experimental data, and are very close to the dielectric constants of their cubic form.
\end{abstract}
\maketitle

\section{Introduction}
In the complementary metal oxide semiconductor (CMOS) devices, SiO$_2$ has been the first and only choice of gate oxide material for a long time.
However, with the continued decrease in the feature size of the CMOS devices, SiO$_2$ is no longer reliable as a gate oxide due to the high tunneling leakage current through it at small thickness. Major research efforts are continuously being carried out in search for a suitable replacement of the SiO$_2$ as a gate material.\cite{Fiorentini} Several promising candidates are metal oxides such as HfO$_2$,\cite{Zhao3, Ceresoli} ZrO$_2$\cite{Zhao2, Zhao1, Ceresoli} and Al$_2$O$_3$,\cite{Mazaleyrat} all of which have high value of dielectric constant $\kappa$. High-$\kappa$ metal oxides in their amorphous (a-) form are more preferable as a gate oxide over their crystalline form due to several important advantages they provide: (i) isotropic physical properties; (ii) no crystalline domain boundary which leads to less defects at the interface with the Si substrate; and (iii) good compatibility with the conventional CMOS fabrication process. Some alloy structures of these high-$\kappa$ metal oxides are also being studied extensively\cite{Scopel,Zhu,Ho,Xiong,Momida,Broqvist} and, in fact, a Hf based alloy material has already been in its third generation of production as a gate oxide in the semiconductor industry and further improvements of thermal stability, dielectric constant and material preparations are underway.\cite{Zhu,Ho,Xiong}

In clear contrary to the abundance of experimental results in the literature, theoretical studies of the structure and dielectric properties of amorphous metal oxides and their alloys are quite limited. One of the reasons is the difficulties in generating reasonable and reliable amorphous structures with the available theoretical methods. Experimentally, the high-$\kappa$ materials are deposited on silicon substrate by vapor deposition \textcolor[rgb]{0,0,0}{ (as-deposited films)} followed by annealing processes at around $1000$ K.\cite{Perevalov, Li} Simulating the deposition process and the resulting amorphous high-$\kappa$ material structures by molecular dynamics (MD) is extremely time consuming because \textcolor[rgb]{0,0,0}{the deposition rate must be controlled close to the experimental value that is usually very low}. The as-deposited a-HfO$_2$ and a-Al$_2$O$_3$ films have been generated by kinetic Monte Carlo (KMC) methods\cite{Dkhissi, Mazaleyrat} with certain assumptions such as the atoms are kept fixed on pre-determined crystal sites and less important dynamics processes of atomic motion are omitted. Both MD and KMC are useful for the simulation of the initial atomic layer deposition processes of the amorphous film growth. To generate bulk amorphous structures, \textit{ab inito} MD simulations can be used in a melt and quench scheme, a-HfO$_2$,\cite{Chen,Scopel,Ceresoli} a-ZrO$_2$\cite{Vanderbilt,Zhao1,Ceresoli} and their silicates\cite{Broqvist,Scopel} have been successfully simulated this way. The activation-relation technique is another method to generate structures of continuous disordered systems\cite{ART} and has been applied to generate a-ZrO$_2$ structures.\cite{Vanderbilt} In this method, one optimizes the structures many times to determine a local energy minima (in configurational space) using a force field which can be calculated from quantum mechanical first principles or described by an empirical potential function. For the former the calculation cost is high while for the latter a reliable potential function is required. Recently, several extended-Tersoff potentials\cite{Shan} were proposed for certain metal oxides and applied to generate their amorphous structures from classical MD.\cite{Kaneta,Momida}

Amorphous structures are long range disordered systems which typically requires much larger than $\sim 100$ atoms to simulate. Another issue, perhaps more difficult to deal with, is the simulation time scale that is needed for the atoms to evolve into the amorphous structure. While \emph{ab initio} methods have simulated up to picoseconds time scale, one certainly wishes to investigate the structures and electronic properties of metal oxides at much larger time scales and much larger sizes in order to reveal the long time limit of the structural evolution and the physical properties of the resulting amorphous material. It is the purpose of this work to report such an investigation.

\begin{figure}
\includegraphics[width=0.9\columnwidth]{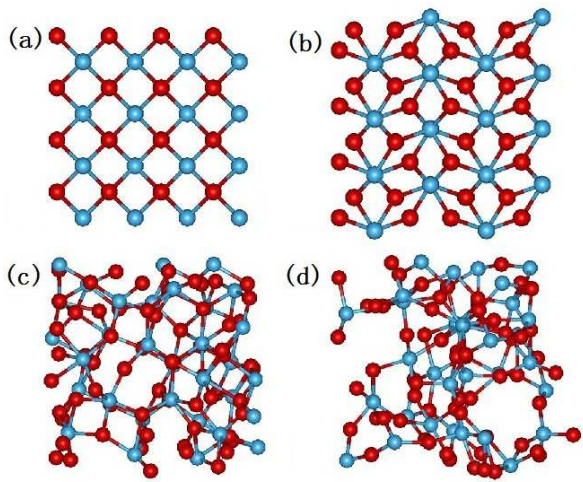}\\
\caption{(color online) Structures of a-HfO$_2$ generated by classical MD starting from the cubic crystal form of HfO$_2$ with different lattice constants, corresponding to different densities. If the density is high, the system recrystallizes to its cubic form (a) or any other crystal form (b); if the density is low, the system tends to form disordered structure with big holes (d), whereas with moderate density of $7.97$ g/cm$^3$ which corresponds to lattice constant of $5.6$ {\AA} for crystalline HfO$_2$, a reasonable amorphous structure is generated (c). Blue spheres represent Hf atoms, red spheres are for O atoms.}
\label{fig1}
\end{figure}

In particular, we calculate the lattice dielectric constants of bulk amorphous metal oxides by a combined approach of classical MD - for structure evolution, and quantum mechanical first principles DFT - for electronic structure analysis. We overcome the size and time scale difficulty by classical MD simulations in a melt and quench scheme and generate amorphous structures of HfO$_2$, ZrO$_2$ and their alloys in the form of Hf$_{1-x}$Zr$_x$O$_2$ with different values of the concentration x ($0<x<1$). The classical MD is based on the interatomic Born-Mayer-Buckingham potential function.\cite{Lewis} This potential has been developed over many years and extensively applied to investigate the growth processes of Y$_2$O$_3$, ZrO$_2$ and Mg-Al-O thin films.\cite{Perumal, Arima, Kilo, Georgieva1, Georgieva2} \textcolor[rgb]{0.00,0.00,0.00}{Here we show that high-$\kappa$ bulk materials of amorphous nature can be generated quite efficiently using this particular potential function.} Detailed analysis of the amorphous structures are carried out including the coordination number, the radial distribution function and the potential energy per atom of the  structures. Using the atomic structures generated this way, we carry out first principles density functional theory (DFT) calculations of the lattice dielectric constant and the results are averaged over ten independent samples for each material. The sample averaged dielectric constants show very good agreement with the corresponding experimental data.

The combined approach of classical MD and DFT is an excellent way to deal with the problem of large sizes and large time scales required by the materials physics of metal oxides. Several advantages can be summarized:
(i) classical MD is inexpensive for simulating systems containing large number of atoms which is important for a proper characterization of the amorphous structures having long range disorders, here we have gone up to 768 atoms although several thousand atoms can be easily handled; (ii) in the melt and quench scheme a much longer relaxation time scale, here to several nano seconds, can be evolved after the material is heated above the melting point. Such a long time scale is important for erasing the memory of the initial structure thereby the final outcome becomes independent of the initial conditions. Here we show that the same kind of amorphous structure is obtained by evolving from different initial structures including from a completely random structure. (iii) Since the calculated physical properties of amorphous material should be averaged over an ensemble of samples due to the randomness of the structure, the combined classical-quantum approach allows such configurational average to be done. Here we show that the calculated dielectric constants of the metal oxides become spatially isotropic after configurational average, but it is typically anisotropic if only one sample is calculated.

The paper is organized as follows. In Sec. II we present the classical MD method as employed in this work. In Sec. III we discuss the structural properties of our simulated amorphous structures, and also present our results for the lattice dielectric constants of these structures. Section IV concludes the paper.

\begin{figure}
\includegraphics[width=0.9\columnwidth]{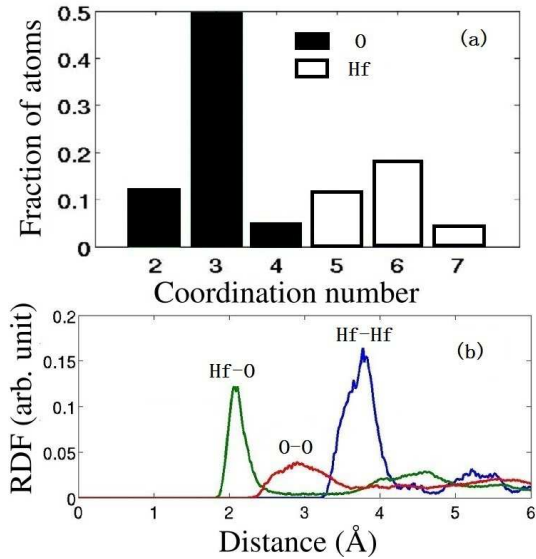}\\
\caption{(color online) (a) The coordination number and (b) the radial distribution function (RDF) of the amorphous HfO$_2$ structure obtained by
the classical MD simulations. Green, red and blue lines in (b) represent Hf-O, O-O and Hf-Hf RDF, respectively. The results are obtained from structures having 768 atoms.}
\label{fig2}
\end{figure}

\section{Methods}

In this section we discuss the potential function used in our classical MD simulations along with a short description of the melt and quench scheme. In our MD method the interactions between the atoms are described by the Born-Mayer-Buckingham potential function\cite{Lewis} as shown below:
\begin{equation}
V(r)=Aexp(-\frac{r_{ij}}{\rho})-\frac{C}{r_{ij}^6}+\frac{q_iq_j}{4\pi\epsilon_0r_{ij}}.
\label{eq1}
\end{equation}
The cation-cation interactions are assumed to be purely coulombic, the parameters for $O^{2-}-O^{2-}$ interactions and cation-anion interactions are presented in Table.~\ref{tab1}. All the parameters are obtained from Ref. [\onlinecite{Lewis}]. We employed Wolf \emph{et. al.}'s method \cite{Wolf, Demonits} for the summation of the Coulomb interaction term.

At the initial stage of the simulation, all the atoms are assigned velocities taken from a Maxwell's distribution at the room temperature $T_a$. The system is then heated up to a temperature $T_m=4000$ K, \textcolor[rgb]{0,0,0}{well above the experimental melting point of around 3000 K,} followed by an evolution of $5$ ns at the same temperature; and then quenched back to the room temperature $T_a$. \textcolor[rgb]{0,0,0}{During this melt and quench process, the motion of an atom is governed by the Langevin equation $m\frac{dv}{dt}=F-\alpha v+F_R$. Actually,} the temperature of the system is adjusted by resetting the velocity of a randomly selected atom at the interval of $\Delta t$ according to the following expression:\cite{Riley}
\begin{equation}
v^{new}=(1-\theta)^{1/2}v^{old}+\theta^{1/2}v^{T_m,T_a}(\xi),
\label{eq2}
\end{equation}
where $\theta$ is a parameter between $0$ and $1$, $v^{T_m, T_a}(\xi)$ is a velocity selected randomly from the Maxwell's distribution at temperature $T_m$ or $T_a$ via a random number $\xi$. \textcolor[rgb]{0,0,0}{Velocity resetting with Eq. (\ref{eq2}) corresponds to taking the damping coefficient $\alpha=m\theta/2\Delta t$ and the random force $F_R=m\theta^{1/2}v^T(\xi)/\Delta t$ in the Langevin equation, where $m$ is the atomic mass.\cite{Riley} Previous calculations \cite{Wang}showed that a value of 0.1 for $\theta$ is reasonable and appropriate.} After the system is quenched to $T_a$, it is allowed to evolve for $10$ ps to calculate the radial distribution function (RDF). Finally, the system is cooled down to $0$ K by a damping method,\cite{Raff} and thus concludes the procedure for generating one amorphous structure of a particular metal oxides.

In the simulations, the initial structure is modeled with a $2\times2\times2$ super cell of Hf$_{1-x}$Zr$_x$O$_2$ containing 96 atoms; or $4\times4\times4$ containing 768 atoms. A periodic boundary condition is imposed on the supercell. The initial atomic configuration is the cubic crystalline form or in the form of a random structure by \textit{randomly} placing the constituents atoms (with the correct ratio) inside the simulation box. These drastically different initial conditions verify that the MD simulation time scales given above are large enough since the same kind of final amorphous structures \textcolor[rgb]{0,0,0}{(similar dielectric constant values and potential energy per atom)} are obtained.

\begin{table}[ht]
\caption{Parameters for Born-Mayer-Buckingham potential function as obtained from Ref. [\onlinecite{Lewis}].}
\centering
\begin{tabular} {ccccc}
\hline\hline
elements & charge & A(eV)  & $\rho$({\AA}) & C \\ \hline
 Hf      & +4      & 1454.6 & 0.350          & 0  \\
 Zr & +4 & 1453.8 & 0.350 & 0 \\
 O  & -2  & 22764.3       &   0.149   & 20.4\\ \hline\hline
\end{tabular}
\label{tab1}
\end{table}

\section{Results}

\subsection{Structure properties of H\lowercase{f}$_{1-x}$Z\lowercase{r}$_x$O$_2$}

To generate reliable amorphous structures it is important to find the optimal density to use for the initial structure. Hence, we first perform a series of independent simulations on the 96-atom super cell, starting from initial structures of different lattice constants (\emph{i.e.} different densities) of cubic HfO$_2$ and ZrO$_2$ in the range from $4.5$ {\AA} to $6.0$ {\AA} with an interval of $0.1$ {\AA}. \textcolor[rgb]{0.00,0.00,0.00}{We found that structures of amorphous nature (see Fig.1c) are obtained when the lattice constant is set to 5.6 A which corresponds to} the density of $7.97$ g/cm$^3$ for HfO$_2$ and $4.66$ g/cm$^3$ for ZrO$_2$. If we start with a lattice constant less than $5.5$ {\AA}, the systems recrystallize to a cubic form (Fig.1a) or some other crystal form (Fig.1b); while with a lattice constant greater than $5.8$ {\AA}, the systems tend to form disordered structures with big holes inside (Fig.1d). Our optimized density values are close to those obtained previously from \textit{ab initio} calculations which were $8.6$ g/cm$^3$ for HfO$_2$ \cite{Kaneta} and $4.86$ g/cm$^3$ for ZrO$_2$.\cite{Vanderbilt} Hence, the lattice constant of 5.6 {\AA} is used as initial condition (cubic form) in subsequent melt and quench evolution to the amorphous structures of Hf$_{1-x}$Zr$_x$O$_2$.

To verify the generated amorphous structures we calculate the coordination numbers and the radial distribution functions. Results are presented in Fig.~\ref{fig2} for a-HfO$_2$ obtained from the 768-atom supercell. From Fig.~\ref{fig2}a we note that the three-coordinated O atoms and six-coordinated Hf atoms dominate the a-HfO$_2$ structures which agree with the known \textit{ab initio} results.\cite{Chen, Scopel} From Fig.~\ref{fig2}b we find that the first peaks in the Hf-O, O-O and Hf-Hf RDFs are at 2.1 {\AA}, 2.8 {\AA}, and 3.6 {\AA}, respectively, again compare well with \textit{ab inito} studies.\cite{Scopel,Kaneta} These comparisons indicates that the classical MD can reliably generate amorphous structures of the metal oxides. Importantly, the efficiency of classical MD allows us to generate ensembles of amorphous samples of Hf$_{1-x}$Zr$_x$O$_2$ for each given concentration $x$, which is essential for statistical analysis of the dielectric properties (see below).

\begin{figure}
\includegraphics[width=0.9\columnwidth]{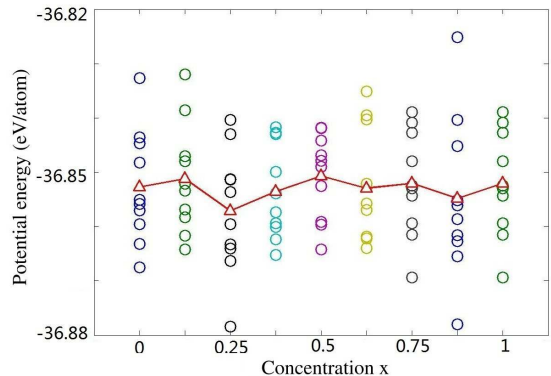}\\
\caption{(color online) Potential energies per atom for the amorphous structures of Hf$_{1-x}$Zr$_x$O$_2$ averaged over $10$ samples for each value of $x$ (solid red line). The potential energies per atom for each individual sample are also shown in the figure (circles). The potential energies of these samples reside in the range of $-36.88$ eV to $-36.82$ eV, which are about 0.3 eV higher than that of the cubic HfO$_2$. The very small differences in the potential energies among the samples with the same value of $x$ confirms the reliability of the structures generated.}
\label{fig3}
\end{figure}

For further analysis, we calculate the potential energy per atom of the a-Hf$_{1-x}$Zr$_x$O$_2$ structures for different values of $x$ ranging from $0$ to $1$ with an interval of $0.125$. For each value of $x$, we generated $10$ different amorphous samples by varying the initial conditions, and then calculated the potential energy per atom for each sample which are presented in Fig.~\ref{fig3}. The results show that the potential energies of a-Hf$_{1-x}$Zr$_x$O$_2$ structures as generated are located in the range from $-36.88$ eV to $-36.82$ eV which is a reasonable and small spread \textcolor[rgb]{0.00,0.00,0.00}{- higher than that of their crystal forms, and have similar potential energies among themselves}. We also generate a structure of a-HfO$_2$ starting from an initial system of $32$ Hf atoms and $64$ O atoms placed \textit{randomly} in a cubic simulation box. The calculated potential energy of this particular a-HfO$_2$ structure comes to $-36.86$ eV which is in \textcolor[rgb]{0,0,0}{the same} energy range of the a-HfO$_2$ structures that were generated from the cubic HfO$_2$ initial conditions. It is very satisfactory that the generated amorphous structures are not dependent on the initial condition, indicating that \textcolor[rgb]{0,0,0}{the simulation time scales} presented above are adequate.

\begin{figure}
\includegraphics[width=0.9\columnwidth]{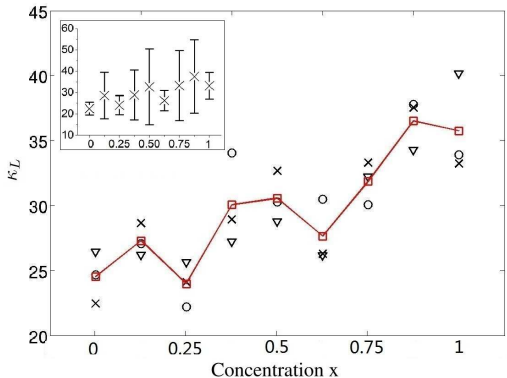}\\
\caption{(color online) Lattice dielectric constants ($\kappa_L$) for the amorphous structures of Hf$_{1-x}$Zr$_x$O$_2$. All the black symbols are for the values in each direction of $X$, $Y$ and $Z$ averaged over $10$ samples, while the square symbols connected with a solid red line represent the values which are averaged over all the three directions for each value of $x$. \textcolor[rgb]{0.00,0.00,0.00}{In the inset the standard deviations of the dielectric constant values over the $10$ samples along the $X$ direction have been presented.} }
\label{fig4}
\end{figure}

\subsection{Dielectric properties of H\lowercase{f}$_{1-x}$Z\lowercase{r}$_x$O$_2$}

A most important property of the high-$\kappa$ metal oxides is their dielectric constant. In this section we present our calculations of the lattice dielectric constant (static) of Hf$_{1-x}$Zr$_x$O$_2$ amorphous structures generated in the last section. We found that the electronic contribution to the dielectric constant is substantially smaller than the lattice contribution and does not change very much for different structures, hence it will not be discussed until when we calculate the total dielectric constant to compare with the measured data (see below).

\begin{table}[ht]
\caption{Calculated lattice dielectric constants for HfO$_2$ and ZrO$_2$ \textcolor[rgb]{0.00,0.00,0.00}{relative to the dielectric constant at vacuum.}}
\centering
\begin{tabular} {ccccc}
\hline\hline
oxide & amorphous & cubic & monoclinic & tetragonal \\ [0.5ex]
\hline
 HfO$_2$ & 24.52 & 31.08 & 14.49 & 125.95 \\
 ZrO$_2$ & 35.76 & 33.80 & 16.14 & 26.72 \\
\hline\hline
\end{tabular}
\label{tab2}
\end{table}

\textcolor[rgb]{0.00,0.00,0.00}{For the calculations of lattice dielectric constants we employ density functional perturbation theory within a projector augmented wave method in the generalized gradient approximation (GGA) using Perdew-Burke-Ernzerhof (PBE) parameterization as implemented in the VASP software.\cite{vasp} The pseudopotentials for Hf and Zr include only the outermost shells in the valence. An energy cutoff of 400 eV is used in the plane wave by sampling the Brillouin zone only at the $\Gamma$ point. The MD-generated amorphous structures have not been relaxed with DFT in VASP. This is reasonable since we generated by MD a statistical ensemble of structures and the values of the dielectric constant have been averaged over the ensemble.} For each amorphous structure the dielectric constant is calculated for an ensemble of $10$ different samples (the 96-atom super cell). The sample averaged values are presented in Table~\ref{tab2} for a-HfO$_2$ and a-ZrO$_2$ along with the dielectric constants for all $3$ of their crystal forms. We notice that the lattice dielectric constants of a-HfO$_2$ and a-ZrO$_2$ are close to that of their cubic form, suggesting that some cubic-like short range order content may dominate the amorphous structures. Indeed, recently experimental measurements\cite{Li} on sputtering samples of a-HfO$_2$ discovered a cubic-like short range coordination in the amorphous microstructure which led to high measured dielectric constant $\kappa \sim 30$. Our calculated lattice $\kappa$ (see Table~\ref{tab2}) is 24.52, adding the electronic contribution which we found to be around 4.4, the total $\kappa \approx 28.9$ which agrees quite well with the measured data. To confirm that the $\kappa$ of a-HfO$_2$ being close to that of the cubic HfO$_2$, is not due to the cubic initial condition of the structure simulation, we have evolved a single a-HfO$_2$ structure from a completely random initial condition (see previous section): its lattice dielectric constant is found to be 21.785, 25.43 and 25.163 in the three spatial directions respectively; its electronic contribution is found to be 4.45. These values are totally consistent with that obtained from the cubic initial conditions.

In comparison to previous theoretical first principles investigations,\cite{Zhao1,Ceresoli,Vanderbilt} our $\kappa$ values are higher. This difference may be explained by noting that in Ref. [\onlinecite{Zhao1, Ceresoli, Vanderbilt}] the amorphous structures were obtained from a relaxed monoclinic structure with cubic lattice vectors which tends to have lower $\kappa$ values. We should mention that direct comparison with experimental measurement is often difficult due to scarcity of data for dielectric constants of amorphous bulk structures.

\textcolor[rgb]{0.00,0.00,0.00}{Note that the contribution of a given mode to the lattice dielectric constant scales as $\tilde{Z}^{*2}_{\lambda}/\omega^2_\lambda$, where $\tilde{Z}^*$ is the mode effective charge and $\omega_\lambda$ is the frequency of the $\lambda$th IR-active phonon mode, a large $\kappa$ will be obtained with the existence of modes with simultaneously high-$\tilde{Z}^*$ and low-$\omega$.\cite{Zhao2, Zhao3} It is easy to understand that cubic HfO$_2$ (ZrO$_2$) with modes of higher-$\tilde{Z}^*$ and lower-$\omega$ has larger $\kappa$ value than monoclinic structure with modes of lower-$\tilde{Z}^*$ and higher-$\omega$. The amorphous structures may have the similar $\tilde{Z}^*$ and $\omega$ values with their cubic form. This consideration should provide a way to explore possible short range structures within the a-Hf$_{1-x}$Zr$_x$O$_2$.}

The ensemble averaged lattice dielectric constants of all the amorphous structures generated for different values of the concentration $x$ are presented in Fig.~\ref{fig4}. We observe a gradual increase (solid red line) in the values of dielectric constant from a-HfO$_2$, through Hf$_{1-x}$Zr$_x$O$_2$, to a-ZrO$_2$. We also note that the spatial resolved dielectric constants in $X$, $Y$ and $Z$ directions (represented by the black symbols in Fig.~\ref{fig4}) are quite \emph{isotropic} which is expected for amorphous structures. It is very important to note that an isotropic dielectric constant is obtained only when the values are averaged over the ensemble of samples of the same structure (same $x$). For a single sample, we always observe large spreads in the values of dielectric constant in different directions. This finding underpins the major advantage of employing a rather inexpensive classical MD to efficiently generate amorphous structures for the statistical average of the physical properties. \textcolor[rgb]{0.00,0.00,0.00}{Note that all the values of dielectric constants presented in this study are relative to the dielectric constant at vacuum.}

\section{Summary}

Using classical MD simulations within a melt and quench scheme, the amorphous structures of HfO$_2$ and ZrO$_2$, along with their alloy structures of the form of Hf$_{1-x}$Zr$_x$O$_2$, have been generated.
A suitable potential function, namely the Born-Mayer-Buckingham potential, has been adopted for these simulations. The calculated coordination numbers and radial distribution functions of the generated structures agree well with the previous studies. The values of the ensemble averaged lattice dielectric constants of these amorphous structures are nearly
isotropic, close to that of their cubic form, and compare well with available experimental data.\cite{Li} On the other hand, a single sample always produces anisotropic $\kappa$ values. Our results indicate that the combined approach of classical MD - for structure evolution, and quantum mechanical first principles DFT - for electronic structure analysis, is a reliable technique for investigating high-$\kappa$ metal oxides.

\section{Acknowledgements}
We would like to thank H.Z. Shao of Fudan University, C. Yin of DEC and J.N. Zhuang of the University of Hong Kong for useful discussions. This work is supported by the University Grant Council (Contract No. AoE/P-04/08) of the Government of HKSAR, the Small Project Funding of HKU (YW), NSERC of Canada (HG), and CIFAR (HG).


\end{document}